\documentclass[reprint,amsmath,amssymb,aps,prl]{revtex4-1}

\usepackage{graphicx}
\usepackage{dcolumn}
\usepackage{bm}
\usepackage{hyperref}
\usepackage{xcolor} 
\usepackage{mathptmx}
\usepackage[most]{tcolorbox}

\begin{document}

\title {
Strong electron-phonon coupling, electron-hole asymmetry, and nonadiabaticity \\
in magic-angle twisted bilayer graphene
}
\author {
  Young Woo Choi and Hyoung Joon Choi
}
\email {
  h.j.choi@yonsei.ac.kr
}
\affiliation {
  Department of Physics, Yonsei University, Seoul 03722, Republic of Korea
}

\date{September 22, 2018}

\begin{abstract}
We report strong electron-phonon coupling in magic-angle twisted bilayer
graphene (MA-TBG) obtained from atomistic description of the system 
including more than 10 000 atoms in the moir\'{e} supercell. Electronic 
structure, phonon spectrum, and electron-phonon coupling strength $\lambda$
are obtained before and after atomic-position relaxation both in  and 
out of plane.  Obtained $\lambda$ is very large for MA-TBG, with 
$\lambda > 1$ near the half-filling energies of the flat bands, while 
it is small ($\lambda \sim 0.1$) for monolayer and unrotated bilayer graphene. 
Significant electron-hole asymmetry occurs in the electronic structure after 
atomic-structure relaxation, so $\lambda$ is much stronger with hole doping 
than electron doping. Obtained electron-phonon coupling is nearly isotropic 
and depends very weakly on electronic band and momentum, indicating that
electron-phonon coupling prefers single-gap $s$-wave superconductivity.
Relevant phonon energies are much larger than electron 
energy scale, going far beyond adiabatic limit. Our results provide
a fundamental understanding of the electron-phonon interaction in MA-TBG, 
highlighting that it can contribute to rich physics of the system.
\end{abstract}

\maketitle
Interplay between the interlayer coupling and the rotational mismatch 
between two graphene layers in bilayer graphene 
results in flattening of Dirac cones 
at certain special twist angles $\theta_M$, called magic angles 
\cite{Magaud:2010,ref2,ref3}. Recently, correlated insulator behavior and 
superconductivity are experimentally observed near the first magic angle 
$\theta_M=1.08^\circ$, demonstrating rich physics induced by the presence 
of the flat bands \cite{Cao:2018a,Cao:2018b}. In this regard, more detailed 
characterizations for the magic-angle twisted bilayer graphene (MA-TBG) 
are attracting great interest
\cite{p1,p2,p3,p4,p5,p6,p7,p8,p10,p11,a1,a2,a4,a5,a7,
a8,a9,a10,a11,a12,a15,a16,a17,a18,a19,a20,a21,a22,a23,a24,
a27,a29,a30,a31,a32,a33,a34,a35,a37,a38,a39,a40,a41,a42,a43,a44}. 

In addition to the exotic electronic properties, 
it has been observed that low-angle bilayer graphene exhibits atomic-scale 
reconstruction \cite{Yoo:2018}. The essential effect of the lattice relaxation 
is such that the area of the AA stacking region becomes smaller, while the AB stacking 
region is larger, and this effect becomes more important as the twist angle gets 
smaller. Also, it is suggested that the lattice relaxation can affect the 
electronic structure, opening superlattice-induced energy gaps 
at the band edges on both electron and hole sides \cite{Nam:2017}. 
Since these gaps are clearly observed in the experiments 
\cite{Cao:2018a,Cao:2016}, it is necessary to consider the lattice relaxation 
when studying TBG in the low-angle regime.

As the electron-phonon coupling strength $\lambda$ in simple monolayer and 
unrotated bilayer graphene is too weak, superconductivity in MA-TBG is 
suspected to be originated from the electron correlation. 
{\em Ab initio} calculations found that $\lambda$ of monolayer and 
unrotated bilayer graphene is less 
than 0.1 near the charge-neutral Fermi level \cite{Park:2008}. 
If $\lambda$ has a similar value in MA-TBG, it cannot
account for the observed superconducting transition temperature $T_c~\sim 1$~K.

However, since $\lambda$ is proportional to the electron density of states,
$\lambda$ of AB-stacked bilayer graphene (AB-BLG), for example, can be as 
large as 0.28 when the Fermi level is tuned to near the van Hove singularity 
points. This suggests that $\lambda$ is likely to be further enhanced in 
low-angle twisted bilayer graphene where the flattening of Dirac cones brings 
large enhancements of the electron density of states. 
Thus, quantitative estimation of $\lambda$ in low-angle twisted bilayer
graphene can provide a valuable insight into the nature of superconductivity.

In this work, we investigate the electron-phonon interaction in MA-TBG with 
atomistic description of the system including more than 10 000 atoms needed
for the moir\'{e} supercell. We use 
a tight-binding approach for electrons and atomic force constants for phonons.
We find that the electron-phonon coupling strength $\lambda$ in MA-TBG is 
almost directly proportional to the electron density of states and becomes 
greater than 1 near the half-filling energies of the flat bands.
It is shown that the lattice relaxations can bring electron-hole asymmetry 
to the electron density of states and, as a result, the hole-side flat bands
have much stronger $\lambda$ than the electron-side. We also find that 
the electron-phonon coupling depends very weakly on the direction and
magnitude of the electronic crystal momentum. We discuss implications of our 
results for superconductivity in MA-TBG.

Although the electron-phonon interaction can be, in principle, obtained 
accurately by self-consistent density functional perturbation theory 
(DFPT), the large number ($\sim$$10^4$) of atoms in the moir\'e supercell is 
a practical barrier making DFPT calculations very difficult to achieve.
In addition, considering correlation effects between electrons in 
atomistic description also requires challenging development due to the 
large number of atoms.
In our present work, we employ a tight-binding approach with one $p$ orbital
per carbon atom and atomic force constants for atomic vibrations
without considering correlation effects between electrons. 
Our results provide a fundamental understanding 
of the electron-phonon interaction in the system obtained 
from atomistic description of electrons and phonons.

A moir\'e supercell of twisted bilayer graphene is constructed by rotating 
each layer of AA-stacked bilayer graphene by $\theta/2$ and $-\theta/2$, 
respectively. The resulting atomic structure has sixfold 
rotation symmetry axis around the $z$ axis, and three twofold rotation symmetry axes 
that are perpendicular to the $z$ axis, which swap two graphene layers as a result.

Preserving the crystal symmetry of nonrelaxed structure, 
we determine the equilibrium atomic positions by minimizing the total energy 
$U$ that is the sum of in-plane strain energy and interlayer binding energy,

\begin{equation}
  \begin{aligned}
    U =\; & \frac{1}{2} \sum_{l=1}^{2} \sum_{p\kappa\alpha,p'\kappa'\beta} 
    C^\text{MLG}_{p\kappa\alpha,p'\kappa'\beta} \; 
    \Delta\tau^{l}_{p\kappa\alpha} \Delta\tau^{l}_{p'\kappa'\beta}\\
    & + \sum_{p\kappa,p'\kappa'} 
    V_\text{KC} ( \bm{\tau}^{1}_{p\kappa}-\bm{\tau}^{2}_{p'\kappa'})~.
  \end{aligned}
  \label{eq:etot}
\end{equation}

Here $\tau^{l}_{p\kappa\alpha}$ is the $\alpha$ ($\alpha=x,y,z$) component 
of the position of the $\kappa$th atom in layer $l$ located at the $p$th 
moir\'e supercell of TBG, 
$\Delta \bm{\tau^{l}}_{p\kappa} = 
\bm{\tau}^{l}_{p\kappa}-\bm{\tilde{\tau}}^{l}_{p\kappa}$
is the deviation from the nonrelaxed position $\bm{\tilde{\tau}}^{l}_{p\kappa}$,
and $C^\text{MLG}_{p\kappa\alpha,p'\kappa'\beta}$ are force 
constants between two atoms in the same layer up to fourth-nearest neighbors,
taken from Ref.~\cite{Wirtz:2004}, 
which are obtained by fitting to the {\em ab initio} phonon dispersion 
calculations of monolayer graphene. The interlayer binding energy is 
calculated using Kolmogorov-Crespi (KC) potential 
$V_\text{KC}$ that depends on interlayer atomic registry \cite{Kolmogorov:2005}.
Without the interlayer binding energy, our total energy 
function has its minimum, by construction, at the atomic positions of the 
rigidly rotated two graphene layers. With the interlayer binding energy, 
the equilibrium atomic positions show that the area of 
AA-stacked regions is shrunk, while AB-stacked regions expanded,
and interlayer distances in AA-stacked regions become larger than 
AB-stacked regions \cite{Yoo:2018,Nam:2017,Tadmor:2017,Tadmor:2018}.

\begin{figure}
    \includegraphics[width=8.8cm]{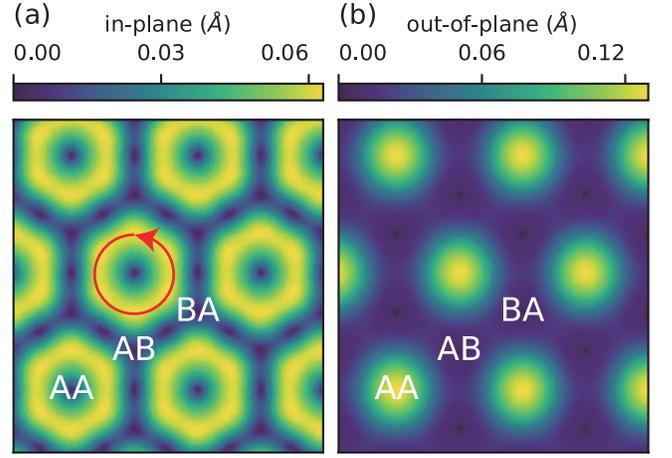}
    \caption {
        Magnitude of (a) in-plane and (b) out-of-plane displacements of 
        the upper layer after the structural relaxation in the TBG at 
        $\theta=1.08^\circ$. Red circular arrow denotes the directions of 
        in-plane displacements.  The other layer has a similar displacement 
        pattern, except that the directions are opposite. The stacking pattern
        of two graphene layers varies within the moir\'e supercell of TBG.
        AA-, AB-, and BA-type stacking regions are denoted by AA, AB, and BA, 
        respectively.
    }
    \label{fig:relax}
\end{figure}

Figure ~\ref{fig:relax} shows the atomic displacements due to the relaxation at 
$\theta=1.08^\circ$. We find that maximal out-of-plane displacements are about 
two times maximal in-plane displacements. Out-of-plane displacements are 
largest at the AA-stacked region, and also noticeable at the AB/BA domain boundary. 
The existence of locally confined strains at AB/BA domain boundaries is one of 
the most important consequences of the lattice relaxations in low angle TBG.
Our results are consistent with previous studies on the lattice relaxations
in low-angle TBG.

To investigate electronic structures of TBG in both nonrelaxed and relaxed 
structure, we employ a single-orbital tight-binding approach where the electronic
Hamiltonian is 
\begin{equation}
  \label{h_tb}
    \hat{H} = \sum_{p\kappa,p'\kappa'} 
    t(\bm{\tau}_{p\kappa}-\bm{\tau}_{p'\kappa'})
    |\phi_\kappa;\bm{R}_p\rangle \langle\phi_{\kappa'};\bm{R}_{p'}|~,
\end{equation}
where $|\phi_\kappa; \bm{R}_p\rangle$ is a carbon $p_z$-like orbital at 
$\bm{\tau}_{p\kappa}$. Here we drop the layer index on $\bm{\tau}_{p\kappa}$,
$\kappa$ sweeps all atoms in both layers, 
and $\bm{\tau}_{p\kappa}=\bm{\tau}_{0\kappa}+\bm{R}_p$
for the $p$th moir\'e supercell at $\bm{R}_p$.
We use the Slater-Koster-type hopping integral,
\begin{eqnarray}
t(\bm{d}) &=& V_{pp\pi}^0 e^{-(d-a_0)/\delta}
	\{ 1-(d_z/d)^2 \} \nonumber \\
    && + \; V_{pp\sigma}^0 e^{-(d-d_0)/\delta}(d_z/d)^2~,
\end{eqnarray}

where $\bm{d}$ is the displacement vector between two orbitals.
The hopping energy $V^0_{pp\pi}=-2.7\,\text{eV}$ is between in-plane nearest
neighbors separated by $a_0=a/\sqrt{3}=1.42\,\text{\AA}$, and
$V^0_{pp\sigma}=0.48\,\text{eV}$ is between two 
veritcally aligned atoms at the distance $d_0 = 3.35\,\text{\AA}$. 
Here $\delta = 0.184 a$ is chosen to set the magnitude of 
the next-nearest-neighbor 
hopping amplitude to be $0.1V^0_{pp\pi}$ \cite{Moon:2012,Ando:2000}. 
We use the cutoff distance 
$d_c = 5\,\text{\AA}$, beyond which the hopping integral is negligible.

\begin{figure}
    \includegraphics[width=9.0cm]{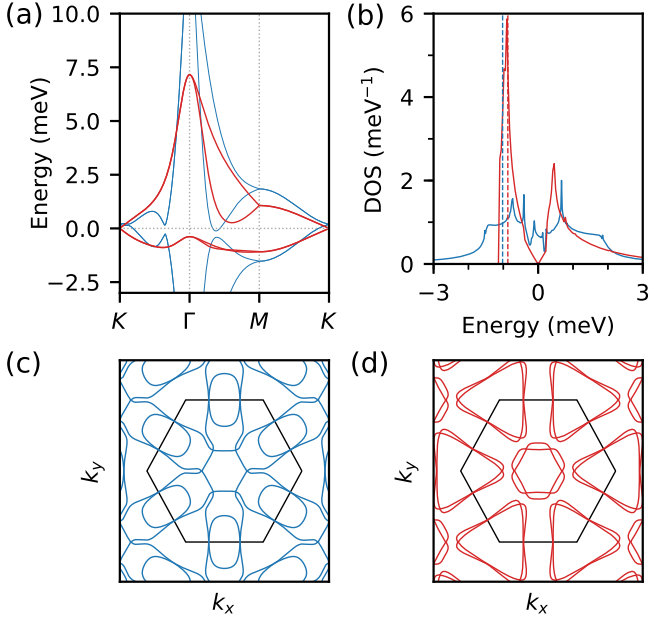}
    \caption {Electronic structure of MA-TBG.
        Tight-binding (a) band structure and 
	(b) density of states per spin per moir\'e supercell for
        nonrelaxed (blue) and relaxed (red) structures at $\theta=1.08^\circ$. 
        Vertical dashed lines show hole-side half-filling energies.
	(c),(d) Fermi surfaces at energies denoted 
	by dashed lines in (b) for nonrelaxed and relaxed structures, 
	respectively.  
    }
    \label{fig:tb}
\end{figure}

Figure \ref{fig:tb}(a) shows the band structures for TBLG at $\theta=1.08^\circ$ 
in the nonrelaxed and relaxed structure. One of the most noticeable effects of 
the lattice relaxation is the opening of the gaps at the edges of the flat 
bands. Furthermore, the electron-side and hole-side flat bands become 
significantly asymmetric due to the relaxation. The hole side gets much 
narrower than the electron side so that the peak height of the density of states 
[Fig.~\ref{fig:tb}(b)] in the hole side is more than twice the electron side. 
The gap opening and the electron-hole asymmetry are consistent 
qualitatively with previous results considering in-plane relaxation 
only \cite{Nam:2017}. 

Figures \ref{fig:tb}(c) and (d) show Fermi surfaces at energies where the 
hole-side flat bands are half-filled for the nonrelaxed and relaxed structures,
respectively. At these energies, 
Fermi surfaces become more complicated than those near the 
charge-neutral energy, where only circular Fermi sheets originating from the 
Dirac cones are located at Brillouin zone corners.
At half-filling energies, Fermi sheets at the zone corners become 
similar to triangles, and the additional $\Gamma$-centered Fermi sheets appear.

Phonons in twisted bilayer graphene are calculated using atomic force constants
$C_{p\kappa\alpha,p'\kappa'\beta} = 
\partial^2 U / \partial \tau_{p\kappa\alpha} \partial \tau_{p' \kappa'\beta}$, 
where $U$ is given by Eq.~(\ref{eq:etot}).
Since we treat the in-plane strain energy  with the harmonic approximation,
the in-plane force constants are unaltered by the lattice relaxation.
The interlayer force constants, however, are evaluated at relaxed atomic
positions because the KC potential \cite{Kolmogorov:2005} is not harmonic. 
Our approach is similar to Ref.~\cite{Cocemasov:2013}, 
except that the Lennard-Jones interlayer potential 
between two graphene layers is replaced 
by the KC potential which can account for registry-dependent energy 
differences in TBG.
From the force constants, we obtain the dynamical matrix 
$D_{\kappa\alpha,\kappa'\beta}(\bm{q})= \sum_{p} 
e^{i\bm{q}\cdot\bm{R}_p} \; C_{0\kappa\alpha, p\kappa'\beta}/M_C $
for phonon wave vector $\bm{q}$,
where $M_C$ is the mass of a carbon atom.
Then, we solve the phonon eigenvalue problems 
$\omega^2_{\bm{q}\nu} \; e_{\bm{q}\nu,\kappa\alpha} = 
\sum_{\kappa',\beta} D_{\kappa\alpha,\kappa'\beta}(\bm{q}) \; 
e_{\bm{q}\nu,\kappa'\beta}$ 
at the irreducible Brillouin zone of TBG for the energy $\omega_{\bm{q}\nu}$
and polarization vector $\bm{e}_{\bm{q}\nu,\kappa}$ of the $\nu$th phonon
mode.
The phonons in the rest of the Brillouin zone are obtained from the symmetry 
relations \cite{Maradudin:1968}. 
We considered all
phonon modes in the moir\'e supercell to obtain unbiased results for electron-phonon interaction.

\begin{figure}
    \includegraphics[width=9.0cm]{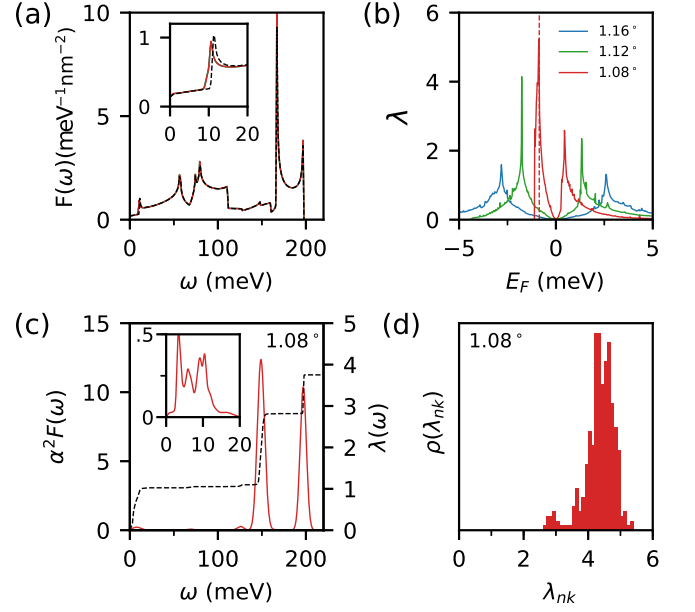}
    \caption{
      (a) Phonon density of states for AB-BLG (dashed black), 
      TBG at $\theta=1.08^\circ$ (solid red), $\theta=1.12^\circ$ (solid green),
      and $\theta=1.16^\circ$ (solid blue). Phonons are insensitive to the 
      small twist-angle differences between those angles.
      The inset shows the frequency range of the interlayer shear and 
      breathing modes, which are softened by the twist.
      (b) Total electron-phonon coupling strength $\lambda$ in TBG
      as a function of the Fermi energy ($E_F$). The vertical red dashed line 
      denotes the energy where hole-side flat bands in $\theta=1.08^\circ$ 
      are half-filled.
      (c) Eliashberg function $\alpha^2F(\omega)$, shown in red, at the 
      half-filling energy in the hole side. The dashed black line denotes 
      $\lambda(\omega)=2\int^{\omega}_0 \alpha^2 F(\omega')/\omega' d\omega'$.
      The inset shows the low-frequency range of $\alpha^2F(\omega)$. 
      Phonon modes at this range contribute to about 30\% of the total 
      coupling strength.
      (d) Distribution of band- and momentum-resolved coupling strength 
      $\lambda_{n\bm{k}}$ of Eq.~(\ref{eq:4a}) 
      at the hole-side half-filling energy.}
    \label{fig:ep}
\end{figure}

Figure ~\ref{fig:ep}(a) shows phonon density of states $F(\omega)$ for 
$\theta=1.08^\circ,1.12^\circ,\mathrm{and}\;1.16^\circ$ as well as AB-BLG. 
Phonon spectra are nearly insensitive to small twist-angle differences. 
So a tiny difference is that, compared to AB-BLG,
interlayer breathing modes near $\omega\sim11\;\mathrm{meV}$ are
slightly softened in TBG. 
[see Fig. S1(a) in the Supplemental Material \cite{SM} for phonon dispersions in AB-BLG].

Now, we calculate the standard electron-phonon coupling strengths
defined as
\begin{subequations}
\begin{eqnarray}
    \label{eq:4a} 
\lambda_{n\bm{k}} &=& 2 N_F \sum_{m\bm{q}\nu} 
  \frac{|g_{mn\nu}(\bm{k},\bm{q})|^2}{\omega_{\bm{q}\nu}} W_{m\bm{k+q}}, \\
  \lambda &=& \sum_{n\bm{k}} \lambda_{n\bm{k}} W_{n\bm{k}},
    \label{eq:4b} 
\end{eqnarray}
\end{subequations}
where $N_F$ is the electron density of states per spin at the Fermi level 
$E_F$, and $W_{n\bm{k}}=\delta(E_F - \varepsilon_{n\bm{k}})/N_{F}$ is the
partial weight of the density of states.
Here, $\varepsilon_{n\bm{k}}$ is the electron energy of the $n$th band with
wavevector $\bm{k}$, and
$W_{n\bm{k}}$ is obtained by the linear tetrahedron method \cite{Bloechl:1994}.
The electron-phonon matrix elements
$g_{mn\nu}(\bm{k},\bm{q})=\langle m\bm{k+q} |\delta_{\bm{q}\nu} \hat{H} 
|n\bm{k}\rangle$ 
couple the electronic states $|n\bm{k}\rangle$ and $|m\bm{k+q}\rangle$,
where $\delta_{\bm{q}\nu} \hat{H}$ is the change in $\hat{H}$ 
due to phonon mode $(\bm{q}\nu)$. 
The electron-phonon matrix elements in localized orbital basis can be expressed
in terms of the changes in the hopping matrix elements due to the atomic 
displacements of phonon modes \cite{Giustino:2007,Gunst:2016,Bernadi:2018},
\begin{eqnarray}
 g_{mn\nu}(\bm{k},\bm{q})
&= & l_{\bm{q}\nu}\sum_{\kappa \alpha} e_{\bm{q}\nu,\kappa \alpha} 
     \sum_{pp',ij}
     e^{-i(\bm{k+q})\cdot\bm{R}_{p'}} e^{i\bm{k}\cdot\bm{R}_p} \nonumber \\
     && \times c^*_{m\bm{k+q},j} c_{n\bm{k},i} 
     \langle\phi_j; \bm{R}_{p'}| 
     \frac{\partial \hat{H}}{\partial \tau_{0 \kappa \alpha}} 
     |\phi_i; \bm{R}_p\rangle,
\end{eqnarray}
where $l_{\bm{q}\nu}=\sqrt{\hbar/(2M_C\omega_{\bm{q}\nu})}$ is the length scale
of phonon mode $(\bm{q}\nu)$, and $c_{n\bm{k},i}$ is the coefficient of the 
electron wavefunctions in local orbital basis, i.e., 
$c_{n\bm{k},i}e^{i\bm{k}\cdot\bm{R}_p}=
\sqrt{N}\langle\phi_i;\bm{R}_p|n\bm{k}\rangle$.
Here, $N$ is the total number of unit cells
over which the electronic states are normalized.
Thus, in our tight-binding approach, 
we obtain the electron-phonon matrix elements 
\begin{eqnarray}
 g_{mn\nu}(\bm{k},\bm{q}) 
&= &  l_{\bm{q}\nu} \sum_{\kappa \alpha}
       e_{\bm{q}\nu,\kappa \alpha} \sum_{p,i} 
       \frac{\partial}{\partial x_\alpha} 
       t(\bm{\tau}_{0\kappa} - \bm{\tau}_{pi}) \nonumber \\
       && \times \{
  e^{i\bm{k}\cdot\bm{R}_p}  c^*_{m\bm{k+q},\kappa}  c_{n\bm{k},i} \\
    && +e^{-i(\bm{k+q})\cdot\bm{R}_p}  c^*_{m\bm{k+q},i} c_{n\bm{k},\kappa} \}.
\end{eqnarray}
When we apply our method to calculate $\lambda$ for simple monolayer graphene and 
AB-BLG, $\lambda$ is less than 0.1 near the charge-neutral energy
but it increases up to $0.2-0.3$ in proportion to the density of states 
when the chemical potential is varied [Fig.~S1(b) \cite{SM}]. 
This is consistent with the previous studies
for monolayer and bilayer graphene \cite{Park:2008,epc1,epc2}.

Figure \ref{fig:ep}(b) shows calculated electron-phonon coupling strength
as a function of the Fermi energy ($E_F$) for the three twist angles.
$\bm{k}$  and $\bm{q}$ grids of $30\times30$ in the moir\'e Brillouin zone
are used for electrons and phonons.
Due to the large density of states of the flat bands, 
$\lambda$ becomes extremely large as $\theta$ approaches $1.08^\circ$,
where the Dirac cones are nearly flat. 
The average interaction between electronic states, $\lambda/N_F$,
for $\theta = 1.08^\circ$ is approximately twice that for $\theta = 0$. 
Furthermore, as the lattice relaxation brings electron-hole asymmetry in the 
density of states,
the maximum value of $\lambda$ in the hole-side flat bands 
is almost twice that in the electron side for $\theta = 1.08^\circ$. 

Figure \ref{fig:ep}(c) shows the isotropic Eliashberg function 
$ \alpha^2F(\omega) = \frac{1}{N_F}\sum_{nm\nu\bm{k}\bm{q}} 
|g_{mn\nu}(\bm{k},\bm{q})|^2 \delta(E_F-\varepsilon_{n\bm{k}}) 
\delta(E_F-\varepsilon_{m\bm{k+q}}) \delta(\omega-\omega_{\bm{q}\nu})$
at the half-filling energy of the 
hole-side flat bands in $\theta=1.08^\circ$. 
With $\alpha^2F(\omega)$, $\lambda$ of Eq.~(\ref{eq:4b}) is equal to
$\lambda=2\int^{\infty}_0 \alpha^2 F(\omega)/\omega d\omega$. 
In Fig.~\ref{fig:ep}(c),
in-plane optical modes generate strong peaks at 150 and 200 meV 
in $\alpha^2F(\omega)$, contributing to about 70\% of $\lambda$.
Although the interlayer shear ($\sim$2 meV) and 
breathing modes ($\sim$11 meV) have an order of magnitude smaller values of 
$\alpha^2F(\omega)$ than the in-plane optical modes,
they have significant contributions to $\lambda$
due to their low-phonon energies. 
We also find $\lambda_{n\bm{k}}$ is nearly isotropic and depends very weakly
on the electronic band and momentum [Fig.~\ref{fig:ep}(d)], 
which indicates the electron-phonon coupling prefers 
single-gap $s$-wave superconductivity \cite{single_gap}.

In conventional phonon-mediated superconductors, 
transition temperature can be reliably calculated from the Migdal-Eliashberg
equations \cite{Migdal:1958,Eliashberg:1960}.
But the validity of the Migdal-Eliashberg theory depends on the existence of 
small parameter $\omega_{\mathrm{ph}}/E_F\ll1$ where $\omega_{\mathrm{ph}}$ is relevant phonon 
energy scale.
This condition is obviously violated in magic-angle twisted bilayer graphene.
For instance, while $E_F \approx 1 \; \mathrm{meV}$ near the half-fillings of
the flat bands, $\omega_{\mathrm{ph}} \approx 2\sim11 \; \mathrm{meV}$ for the 
interlayer shear and breathing modes, and 
$\omega_{\mathrm{ph}} \approx 150\sim200 \; \mathrm{meV}$ for the in-plane optical modes.
In this sense, MA-TBG systems are close to the antiadiabatic limit 
$\omega_{\mathrm{ph}}/E_F\gg1$.

In the antiadiabatic limit, $T_c$ was studied in several literatures 
\cite{Eagles:1967,Ikeda:1992,Gorkov:2016a,Gorkov:2016b,Sadoskii:2018},
where the prefactor of $T_c$ is determined by $E_F$ instead of the phonon energy 
\cite{Gorkov:2016a,Sadoskii:2018}, that is, 
\begin{equation}
  T_c\sim E_F \exp(-1/\lambda).
\end{equation}
In our calculations for $\theta=1.08^\circ$, $\lambda$ = 3.6 (0.56) at 
$E_F$ = 0.86 (1.02) meV when the hole-side (electron-side)
flat bands are half-filled.
These values give $T_c\sim 7.5\;\mathrm{K}$ for the hole side and
$T_c \sim 1.9\;\mathrm{K}$ for the electron side.
Although our estimation is crude for direct comparison with experiments,
the order of magnitude is close to the experimentally observed 
$T_c\sim1.7\;\mathrm{K}$ in the hole side.
Since our estimation did not consider
the effect of Coulomb interaction, which can reduce $T_c$, 
we expect that calculating $T_c$ including the Coulomb effect 
can give more consistent results to the experimental situations.
Also, the rapid energy dependence of the electronic density of states can
play an important role in determining $T_c$ \cite{Allen:1982}.

In conclusion, we have calculated the electron-phonon coupling 
strength 
in the magic-angle twisted bilayer graphene using atomistic description 
of electrons and phonons. Obtained $\lambda$ in MA-TBG becomes almost 
an order of magnitude larger than that in simple monolayer or unrotated 
bilayer graphene. For $\theta=1.08^\circ$, the electron-hole asymmetry 
arises from atomic-structure relaxation due to interlayer interaction so that 
the electron-phonon coupling is stronger in the hole-side flat bands.
The obtained electron-phonon interaction is almost isotropic and depends 
very weakly on the electronic band and momentum, 
which indicates the electron-phonon coupling
prefers single-gap $s$-wave superconductivity.
We also found that MA-TBG is in the antiadiabatic limit
where the electron energy scale is much smaller than the phonon energy scale.
Although the $T_c$ formula in the antiadiabatic limit produces values 
of $T_c$ comparable to the experiments, theory of $T_c$ of the system
may require including Coulomb interaction and rapid energy dependence of the 
electronic density of states as well as electron correlation and any possible 
presence of magnetic fluctuations. Our results provide a fundamental 
understanding of the electron-phonon interaction in MA-TBG obtained from an atomistic
description of electrons and phonons, highlighting that it can contribute 
to rich physics of the system.

\begin{acknowledgments}
This work was supported by National Research Foundation of Korea (Grant No.~2011-0018306). 
Y.W.C. acknowledges support from National Research Foundation of Korea 
(Global Ph.D. Fellowship Program NRF-2017H1A2A1042152).
Computational resources have been provided by KISTI
Supercomputing Center (Projects No.~KSC-2017-C3-0079).
\end{acknowledgments}


\pagebreak

\onecolumngrid
\setcounter{equation}{0}
\setcounter{figure}{0}
\setcounter{table}{0}
\renewcommand{\theequation}{S\arabic{equation}}
\renewcommand{\thefigure}{S\arabic{figure}}
\renewcommand{\bibnumfmt}[1]{[S#1]}
\renewcommand{\citenumfont}[1]{S#1}

\begin{center}
  \textbf{\large Supplemental Material:\\
Strong electron-phonon coupling, electron-hole asymmetry, and 
    nonadiabaticity \\in magic-angle twisted bilayer graphene}\\[.2cm]
  Young Woo Choi and Hyoung Joon Choi$^*$\\[.1cm]
  {\itshape Department of Physics, Yonsei University, Seoul 03722, Republic of Korea\\}
(Dated: September 22, 2018)\\[1cm]
\end{center}

\begin{center}
\setlength{\fboxrule}{0pt}

\fbox{\begin{minipage}{0.8\textwidth}
    \hspace{5pt} This supplemental material provides (i) phonon dispersions of unrotated bilayer graphene obtained by our method and (ii) the electron-phonon interaction strengths of monolayer and unrotated bilayer graphene obtained by our method.  
\end{minipage}}
\end{center}

\vspace{1cm}

Figure S1(a) shows that the phonon dispersions of AB-stacked bilayer graphene 
obtained by our method, which are in good agreement with those 
from {\em ab initio} density functional perturbation theory (DFPT). 
Figure S1(b) shows the electron-phonon coupling strengths for monolayer 
graphene and AB-stacked bilayer graphene
obtained by our method.

\begin{figure*}[h]
    \includegraphics[width=13cm]{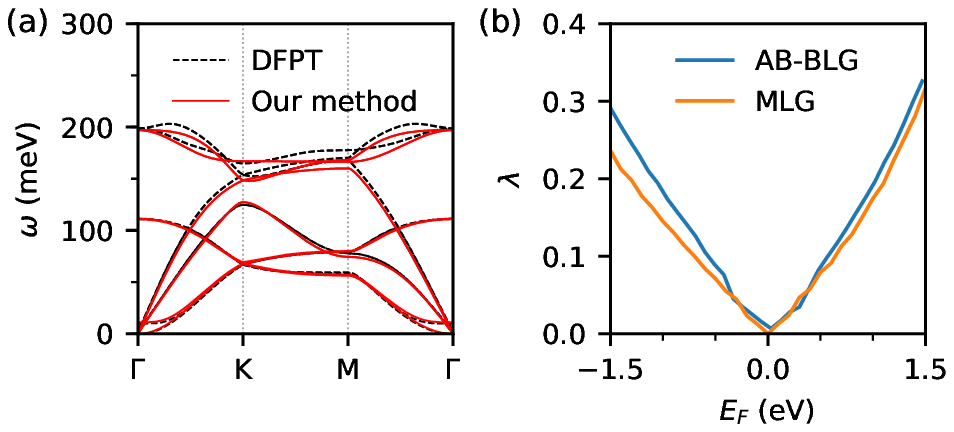}
    \caption {
(a) Phonon dispersions of AB-stacked bilayer graphene obtained by our method,
shown in red solid lines, and those from {\em ab initio}
density functional perturbation theory (DFPT), shown in black dashed lines.
Phonon dispersions are plotted along the high-symmetry lines 
in the two-dimensional Brillouin zone of unrotated bilayer graphene.
(b) The electron-phonon coupling strengths for monolayer graphene (MLG) 
and AB-stacked bilayer graphene (AB-BLG) obtained by our method, as functions 
of the Fermi energy ($E_F$).  }
\end{figure*}

\noindent $^*$ h.j.choi@yonsei.ac.kr

\begin{thebibliography}{99}
\bibitem{single_gap} 
  If distribution of $\lambda_{n\bm{k}}$ formed multiple groups in 
  Fig.~\ref{fig:ep}(d), it would indicate multiple-gap $s$-wave 
  superconductivity as in MgB$_2$. However, in Fig.~\ref{fig:ep}(d), 
  distribution of $\lambda_{n\bm{k}}$ forms a single group centered at 4.2. 
  This indicates that the electron-phonon interaction in MA-TBG prefers 
  a single uniform superconducting energy gap.
\end{thebibliography}
\end{document}